\documentclass[journal=jpclcd,manuscript=letter,layout=two-column]{achemso}

\usepackage[version=3]{mhchem} 

\usepackage{physics}
\usepackage[T1]{fontenc}
\usepackage{multirow}
\usepackage{tabularx}
\usepackage{xcolor}
\usepackage{makecell}
\usepackage{threeparttable}

\def\mic{CH$_{2}$NH$_{2}^{+}$}

\def\be{\begin{equation}}
\def\ee{\end{equation}}
\def\spci{$\sigma\pi^{\ast}$/S$_0$}
\def\ppci{$\pi\pi^{\ast}$/S$_0$}

\author{Daniil N. Chistikov}
\affiliation[MSU]{Department of Chemistry, Lomonosov Moscow State University, Leninskie Gory 1/3, 119991 Moscow, Russia}

\altaffiliation{Institute of Quantum Physics, Irkutsk National Research Technical University, 83 Lermontov Street, Irkutsk 664074, Russia}

\author{Pavel M. Radzikovitsky}
\affiliation[MSU]{Department of Chemistry, Lomonosov Moscow State University, Leninskie Gory 1/3, 119991 Moscow, Russia}

\author{Dmitry S. Popov}
\affiliation[MSU]{Department of Chemistry, Lomonosov Moscow State University, Leninskie Gory 1/3, 119991 Moscow, Russia}

\author{Ivan V. Dudakov}
\affiliation[MSU]{Department of Chemistry, Lomonosov Moscow State University, Leninskie Gory 1/3, 119991 Moscow, Russia}
\alsoaffiliation{MSU Institute for Artificial Intelligence, Lomonosov Moscow State University, Moscow 119192, Russia}

\author{Vadim V. Korolev}
\affiliation[MSU]{Department of Chemistry, Lomonosov Moscow State University, Leninskie Gory 1/3, 119991 Moscow, Russia}
\alsoaffiliation{MSU Institute for Artificial Intelligence, Lomonosov Moscow State University, Moscow 119192, Russia}

\author{Vladimir E. Bochenkov}
\affiliation[MSU]{Department of Chemistry, Lomonosov Moscow State University, Leninskie Gory 1/3, 119991 Moscow, Russia}

\author{Anastasia V. Bochenkova}
\email{bochenkova@phys.chem.msu.ru}
\affiliation[MSU]{Department of Chemistry, Lomonosov Moscow State University, Leninskie Gory 1/3, 119991 Moscow, Russia}

\title{Machine Learning Photodynamics Unveils a Controlled H$_2$ Loss Channel in Methaniminium Cation}


\begin{document}

\begin{tocentry}
\centering
\vspace{16pt}
\includegraphics[width=5cm]{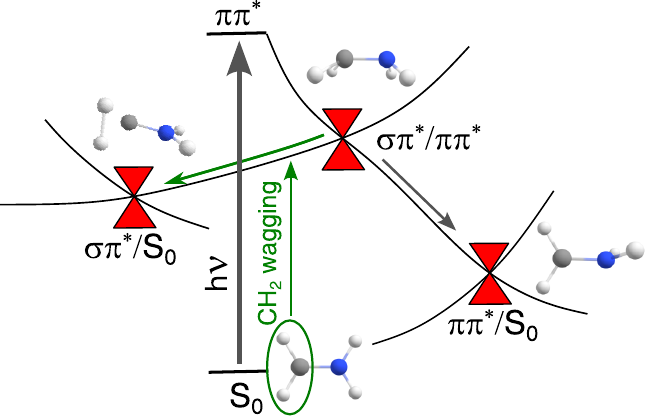}
\end{tocentry}

\begin{abstract}
    The methaniminium cation, CH$_2$NH$_2^+$, plays an important role in Titan's N$_2$--CH$_4$ atmospheric chemistry. As the simplest protonated Schiff base (PSB), it also serves as a model for studying the nonadiabatic dynamics of retinal PSB, the chromophore central to vertebrate vision. While previous studies have established CN bond cleavage and photoisomerization as the primary pathways in the photochemistry of CH$_2$NH$_2^+$, we now report a new UV-induced photochemical pathway to HCNH$^+$, the dominant ion in Titan's upper atmosphere. Through high-level XMCQDPT2 and CASSCF(12,12) calculations, we identify a novel S$_1$/S$_0$ conical intersection that mediates the concerted double H-atom elimination from the carbon center of CH$_2$NH$_2^+$, yielding carbene CNH$_2^+$ as a direct precursor to HCNH$^+$. On-the-fly trajectory surface hopping dynamics confirm the presence of direct H$_2$ loss following excitation to either the S$_2$ or S$_1$ state. Furthermore, our large-scale, machine learning-accelerated simulations reveal that mode-specific pre-excitation can selectively funnel the dynamics into this new channel via the vibronically allowed S$_1$ state, enabling targeted control of the photochemical outcome.
\end{abstract}

Photochemically induced processes play a crucial role in various fields of biology, physics, and chemistry. Study of these processes provides access to the understanding of molecular mechanisms of primary events in vision \cite{polli2010conical,warshel1976bicycle}. Photochemical reactions are key components of the modern organic synthesis \cite{karkas2016photochemical, hoffmann2008photochemical}, their applications also include energy conversion, design of new materials and photovoltaics \cite{long2017nonadiabatic, zhang2013photochromic, liu1990photoelectrochemical}. Since photochemical reactions in organic substances involve many short-lived intermediates on a femto- to picosecond timescale, and thus are difficult to study experimentally, non-adiabatic molecular dynamics (NAMD) is an invaluable tool for investigating such ultrafast photoinduced processes~\cite{Martinez2018_review}. The NAMD studies yield valuable insights into the mechanisms, timescales, and quantum yields of photochemical transformations, while also guiding the design of new experiments and helping to interpret time-resolved pump-probe spectroscopy results.

State-of-the-art NAMD simulations employ fully quantum-mechanical wave packet simulation methods, such as the multi-configuration time-dependent Hartree method~\cite{Worth2003}. Such methods scale exponentially with system size, which severely limits their application. The \textit{ab initio} multiple spawning (AIMS) method has emerged as a powerful alternative that retains a full quantum treatment of key degrees of freedom while enabling on-the-fly dynamics~\cite{Martinez1996,Martinez2000}. However, the prohibitive computational cost of AIMS restricts its application to large systems, precluding its use for high-throughput screening and exhaustive statistical sampling. 

The time-dependent evolution of coupled nuclear and electronic motion is typically described using mixed quantum-classical dynamics, where electronic degrees of freedom are treated quantum mechanically while nuclei are propagated on an excited potential energy surface classically~\cite{Tully1998,Barbatti2018}. The classical treatment of nuclei implies a local approximation, and hence energies and forces acting on them can be computed on-the-fly, without the need to precompute potential energy surfaces. Treating nuclei classically also allows one to run simulations in full dimension. Trajectory surface hopping (TSH) simulates nonadiabatic dynamics by allowing trajectories to stochastically hop between electronic states, with hopping probabilities calculated from the nonadiabatic coupling vectors. The average over a swarm of TSH trajectories provides the description of the nonadiabatic dynamics, capturing the time-dependent populations of electronic states and key molecular motions. While the AIMS method offers superior formal accuracy by explicitly treating quantum decoherence, trajectory-based surface hopping with decoherence corrections remains indispensable for simulating molecular systems due to its vastly lower computational cost and straightforward interpretation~\cite{Subotnik2016}. Despite its simplified treatment of decoherence, TSH produces ensemble-averaged results that are in close agreement with the quantum-mechanical benchmark of AIMS~\cite{AIMS_TSH_D}.

Conical intersections are recognized as key mediators of ultrafast photoinduced dynamics in organic and biological chromophores~\cite{polli2010conical,Gozem2017,BOCHENKOVA2024141}. Targeted control of the dynamics can be enabled by selectively exciting the vibrational modes that mediate nonadiabatic transitions. By depositing energy into these modes, one can fundamentally alter the photochemical pathway. While the principle of vibrational pre-excitation has long been explored in photodissociation studies to achieve selective bond cleavage~\cite{Bar2001,Worth2017}, its application becomes profoundly more complex when multiple electronic states and conical intersections are involved. In such intricate systems, the capacity for targeted control remains a central question. Consequently, high-level nonadiabatic dynamics simulations supported by advanced analysis are indispensable for determining if and how photoinduced dynamics can be steered by selective vibrational excitation.

Exhaustive sampling of excited-state decay pathways requires numerous trajectory runs, rendering on-the-fly nonadiabatic dynamics prohibitively expensive, particularly with high-level multiconfiguration electronic structure theory methods. In recent years, significant efforts have been made to construct excited-state potential energy surfaces using machine learning approaches, aiming to drastically reduce the computational cost of NAMD simulations~\cite{Dral2018,Lan2018,Marquetand2021,Dral2024,Slavicek2024,Jorg2022}. Furthermore, these machine-learned potentials enable the development of advanced tools for analyzing branching ratios and quantifying the mode-specificity of various reaction channels.

The CH$_2$NH$_2^+$ molecule has been a long-standing subject of research interest. It served as a key prototype, alongside other simple organic cations, for studying the ground-state dehydrogenation reaction, specifically the molecular loss of H$_2$~\cite{hoon2001theoretical,suarez1997ab}. It later emerged as both a standard benchmark for NAMD methods and the minimal model for the retinal protonated Schiff base, the chromophore central to vertebrate vision~\cite{barbatti2007fly,tapavicza2007trajectory,yamazaki2005locating,suchan2020pragmatic,pittner2009optimization,west2014nonadiabatic, barbatti2006ultrafast, fabiano2008approximate, westermayr2019machine, westermayr2020combining,lan_jade}. 
Following excitation to the S$_2$ ($\pi\pi^{\ast}$) state, the molecule decays rapidly to the S$_1$ ($\sigma\pi^{\ast}$) state through a conical intersection driven by CN bond elongation.
After switching to the S$_1$ state, the dynamics becomes more complicated. The trajectories evolving on the S$_1$ potential energy surface diverge into at least two distinct reaction pathways~\cite{barbatti2008nonadiabatic}. In 30--50\% of the trajectories, a significant stretching of the CN bond leads to the formation of a weakly bound [CH$_2\cdots$NH$_2$]$^+$ complex, characterized by a CN distance exceeding 2.0~\AA~\cite{barbatti2007fly,fabiano2008approximate}. The remaining trajectories decay to the S$_0$ state via a conical intersection accessed by torsion around the CN bond, with an angle of $\sim$90$^\circ$.
The observed photodissociation channels for CH$_2$NH$_2^+$ yield CH$_2^+$ $+$ NH$_2$ and H-atom elimination products~\cite{barbatti2006ultrafast}.

The significance of the CH$_2$NH$_2^+$ molecule extends beyond its role as a convenient photobiological model; it is also a known intermediate in the nitrogen-carbon reaction chain of Titan's ionosphere~\cite{Nixon2024}. The first \textit{in situ} measurements of Titan's ionosphere, conducted by the Cassini-Huygens mission's Ion and Neutral Mass Spectrometer (INMS), have revealed HCNH$^+$ as the most abundant ion in the upper atmosphere~\cite{singh2010photodissociation}. This cation is produced from CH$_2$NH$_2^+$ by losing either a hydrogen molecule or two hydrogen atoms \cite{pei2012ion,thackston2018quantum}. The CH$_2$NH$_2^+$ cation can be formed through various pathways, including the protonation of CH$_2$NH \cite{vuitton2006nitrogen}, hydrogen loss from methylaminyl cation CH$_3$NH$_2^+$ \cite{singh2010photodissociation}, and the reaction of CH$_4$ with N$^+$($^3$P) \cite{freindorf2021formation}. The study of methanimine CH$_2$NH and its charged species attracts a lot of attention due to their potential role in the formation of tholins (organic aerosols) in atmospheres of icy planets and the interstellar medium \cite{Cable2012,skouteris2015dimerization,freindorf2021formation}. These molecules are regarded as prebiotic precursors, facilitating the formation of amino acids, nucleic acids, and other essential biomolecules.

Here, we revisit the photoinduced non-adiabatic dynamics of CH$_2$NH$_2^+$, revealing a new pathway for its barrierless photodissociation into HCNH$^+$. Using high-level multireference quasi-degenerate perturbation theory XMCQDPT2 theory~\cite{xmcqdpt2}, we disclose a novel type of the S$_1$/S$_0$ conical intersection that mediates the simultaneous loss of two hydrogen atoms from the carbon center. On-the-fly mixed quantum-classical nonadiabatic dynamics simulations based on the SA(3)-CASSCF(12,12) potentials within the newly developed JADE~\cite{lan_jade}/Firefly~\cite{Firefly} interface confirm the significance of this pathway. By constructing neural network potentials and performing exhaustive NAMD simulations, we demonstrate that pre-exciting the CH$_2$ wagging mode selectively steers the photoinduced dynamics through the newly discovered channel, establishing a means to control the photochemical outcome.

\begin{figure}[t!]
\centering
\includegraphics[width=0.5\textwidth]{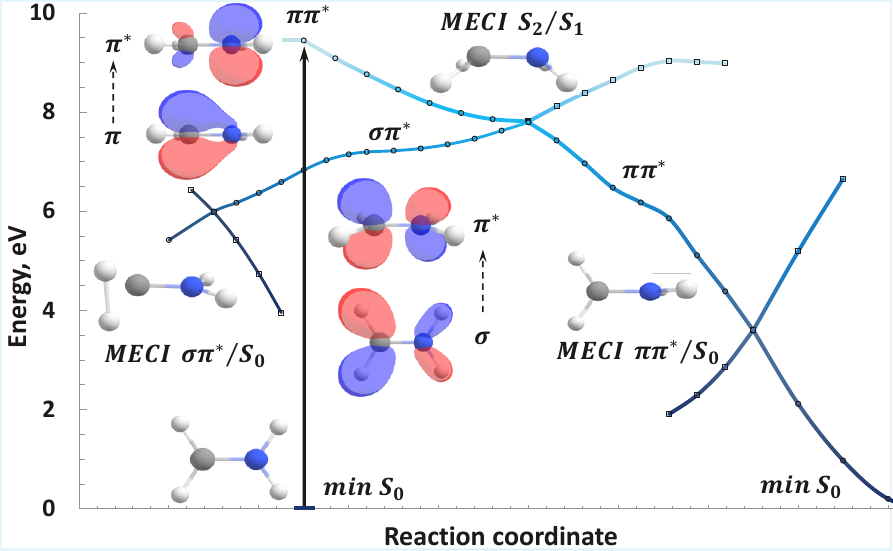}
\caption{Excited-state decay pathways of \mic{}, showing XMCQDPT2 energy scans from geodesic interpolation~\cite{zhu2019geodesic} between the S$_2$ Franck-Condon point and three minimum-energy conical intersections. The natural orbitals of the valence $\pi\pi^*$ and $\sigma\pi^*$ states, primarily involved in the transitions, illustrate the change in electronic character.}
\label{fig:diag}
\end{figure}

\begin{figure*}[t!]
\centering
\includegraphics[width=1.0\textwidth]{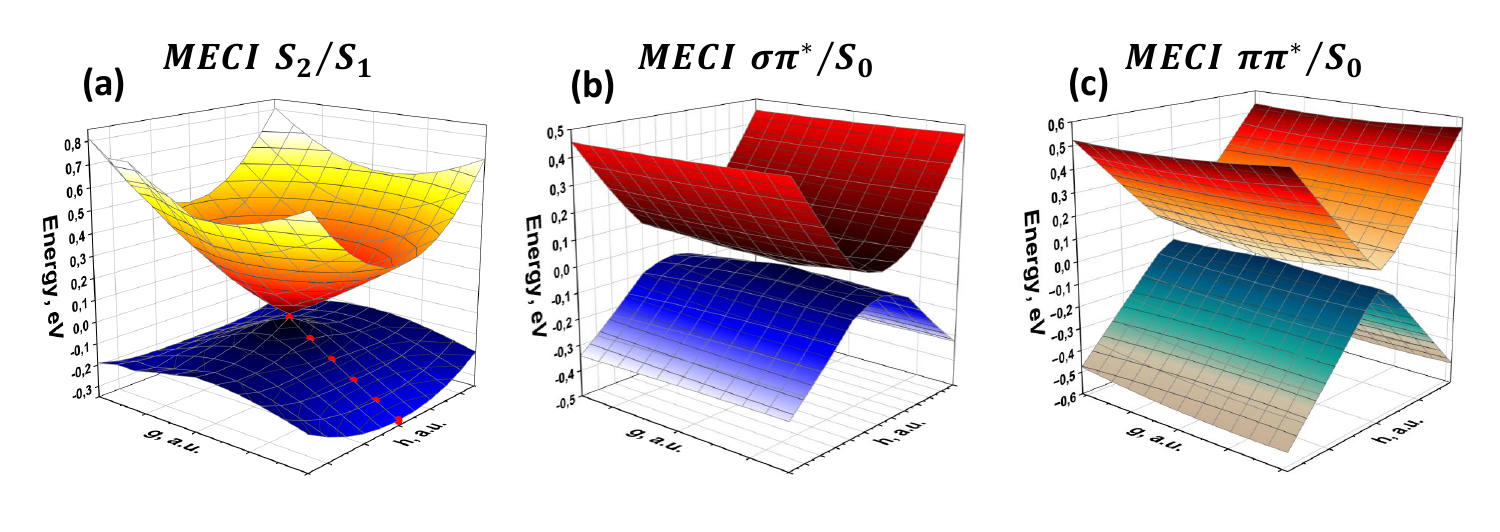}
\caption{XMCQDPT2 topographies of the potential energy surfaces around the minimum-energy conical intersections: (a) S$_2$($\pi\pi^*$)/S$_1$($\sigma\pi^*$) ; (b) S$_1$($\sigma\pi^*$)/S$_0$; (c) S$_1$($\pi\pi^*$)/S$_0$. The surfaces are plotted in the corresponding branching planes~\cite{book:CI} spanned by the gradient difference (\textbf{g}) and nonadiabatic coupling (\textbf{h}) vectors. 
}
\label{fig:topographies}
\end{figure*}

In order to simulate the time evolution of the photoexcited methaniminium cation \mic, we employed two commonly used approaches for simulating semiclassical nonadiabatic dynamics: Tully's fewest switches surface hopping (FSSH)~\cite{tully1990molecular} with an energy-based decoherence correction~\cite{Granucci2010} and the Belyaev-Lebedev Landau-Zener method (LZBL)~\cite{PhysRevA.84.014701}.
The FSSH dynamics were carried out with JADE~\cite{lan_jade}, interfaced with the Firefly quantum chemistry package \cite{Firefly} to compute electronic energies, energy gradients, and wavefunction overlaps. All the dynamics simulations were carried out using the SA(3)-CASSCF(12,12) method and aug-cc-pVDZ basis set. The state-averaging (SA) procedure included three lowest singlet states, which corresponded to the ground and excited ($\sigma\pi^*$, $\pi\pi^*$) electronic states. All core and valence orbitals, except for the 1$s$ orbitals of the C and N atoms, were included in the active space. To compute the overlap between the wavefunctions, used to approximate the scalar product of the nuclear velocity and nonadiabatic coupling (NAC) vectors, we implemented a straightforward procedure using the L\"owdin's formula~\cite{pittner2009optimization}. 

An advantage of the LZBL method is that it does not require calculations of nonadiabatic couplings for the dynamics simulations. This feature enables an efficient machine-learning acceleration of the dynamics, as the FSSH method would otherwise require fitting the NAC surfaces. 
Despite significant efforts~\cite{westermayr2019machine,westermayr2020combining,varga2018direct,li2021new,richardson2023machine}, constructing reliable fitted surfaces for nonadiabatic couplings remains a major challenge. At the same time, the LZBL method is capable of producing the results similar to those obtained by the FSSH method\cite{suchan2020pragmatic}. The LZBL dynamics were performed using the MLatom package\cite{MLatom_paper}, which features an interface for dynamics on MACE~\cite{batatia2022mace} neural network potentials.

In the Franck-Condon region, the S$_2$ state has a $\pi\pi^*$ character, while S$_1$ corresponds to the $\sigma\pi^*$ excited state (Fig.~\ref{fig:diag} and Fig.~S1). Following UV excitation, the photoinduced dynamics primarily begins in the optically bright S$_2$ state. The transition to the S$_1$ state, while formally electronically forbidden at the equilibrium geometry, gains intensity through vibronic couplings. This state can be populated by simultaneously exciting those vibrational modes (\textit{e.g.} out-of-plane vibrations) that break the molecular symmetry from C$_{2v}$ to C$_s$. Therefore, both the S$_2$ and S$_1$ states can be excited in the UV.

Our XMCQDPT2/SA(3)-CASSCF(12,12)/cc-pVDZ calculations reveal three low-lying minimum-energy conical intersections (MECIs) interconnecting the S$_2$, S$_1$, and S$_0$ states (Fig.~\ref{fig:topographies} and Figs.~S2-S5). When dynamics originate in the S$_2$ state, the dominant relaxation pathway involves successive passages through two conical intersections: first, the S$_2$/S$_1$ intersection associated with bipyramidalization and CN bond elongation, and subsequently, the $\pi\pi^*$/S$_0$ intersection associated with torsion around the CN bond and its stretching (Fig.~\ref{fig:diag} and Figs.~S6-S8). This pathway results in barrierless isomerization. The initial excitation to S$_2$ also promotes CN bond dissociation, as the CN stretching mode is highly active in the Franck-Condon region (Fig.~S9). Its excitation directly channels the system into the peaked S$_2$/S$_1$ conical intersection (Fig.~\ref{fig:topographies}a), where the S$_2$ local minimum coincides with the MECI. The resultant nuclear momentum then drives CN bond dissociation upon passage through the $\pi\pi^*$/S$_0$ intersection seam. These findings are consistent with the previously established internal conversion mechanism for \mic~\cite{barbatti2006ultrafast,barbatti2008nonadiabatic,fabiano2008approximate}.

We also reveal a new relaxation pathway through S$_2$/S$_1$ and S$_1$/S$_0$ conical intersections; however, the second conical intersection is of a different nature compared to that of the main relaxation pathway (Fig.~\ref{fig:diag} and Figs.~S6-S8). Unlike the S$_1$ state of the main branch, which remains its $\pi\pi^*$ origin after the passage of the first S$_2$/S$_1$ conical intersection, S$_1$ acquires the $\sigma\pi^*$ character along the minor pathway. At the novel \spci{} conical intersection, the branching plane is spanned by a nonadiabatic coupling vector, involving CH$_2$ asymmetric stretching and rocking, and a gradient difference vector, involving CH$_2$ wagging, scissoring, and symmetric stretching (see animations in the SI). The minor branch leads to photoinduced fragmentation of \mic{} through direct loss of H$_2$. The dissociative character of the $\sigma\pi^*$ state can also drive single H-atom elimination, predominantly from the carbon center.

When dynamics originate in the S$_1$ state, Franck-Condon active modes, specifically the CH$_2$ scissoring and symmetric stretching (Fig. S9), facilitate passage through a novel \spci{} MECI, directing the system to approach the intersection seam along the gradient difference vector. The out-of-plane CH$_2$ wagging, is, however, not directly excited in the Franck-Condon region. The resulting branching ratio between relaxation pathways is therefore determined by the topographies of the multidimensional potential energy surfaces leading to the \spci{} and \ppci{} intersections (Fig. S6). These pathways comprise direct H$_2$ loss via the \spci{} MECI and photoisomerization via the \ppci{} MECI. The dissociative $\sigma\pi^*$ character of S$_1$ still permits single H-atom loss, but the CN dissociation is greatly suppressed compared to the S$_2$ excitation, as the CN stretching mode is no longer Franck-Condon active. Finally, transitions to S$_1$ becomes vibronically allowed through A$_2$, B$_1$, and B$_2$ modes (Fig.~S1), suggesting that vibrational pre-excitation can actively promote this transition.

\begin{figure*}[t!]
\centering
\includegraphics[width=0.9\textwidth]{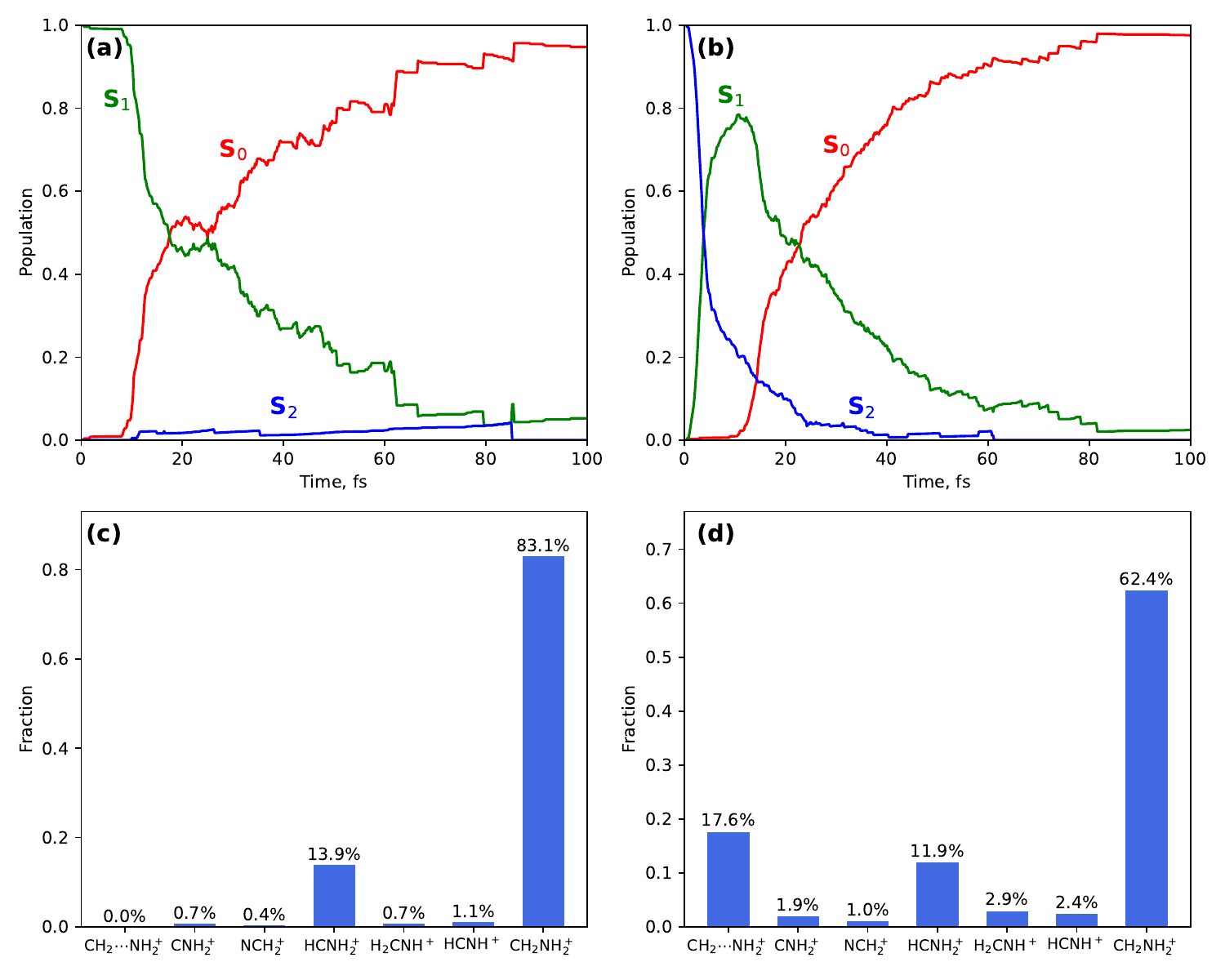}
\caption{CASSCF/FSSH nonadiabatic dynamics following photoexcitation at 0~K. Shown are the electronic state populations over time for excitation to S$_1$ (a) and S$_2$ (b) and the branching ratios of the resulting photodissociation channels starting from S$_1$ (c) and S$_2$ (d).} \label{fig:jade_dynamics}
\end{figure*}

To determine excited-state lifetimes, analyze decay pathways, and elucidate the role of the newly discovered MECI, we performed nonadiabatic dynamics simulations initiated from both the S$_2$ and S$_1$ states. For \textit{ab initio} runs, we selected the SA(3)-CASSCF(12,12)/aug-cc-pVDZ method, as benchmark calculations showed that the more affordable CASSCF approach provides potential energy surfaces and conical intersection geometries similar to those from the higher-level XMCQDPT2 method. In particular, the root-mean-square deviations between the MECI structures are below 0.07~\AA~(Figs. S2–S4), the conical intersections are of the same type at the two levels of theory (Fig.~\ref{fig:topographies} and Fig.~S5), the potential energy scans along the competing branches in S$_1$ leading to the \spci{} and \ppci{} MECIs do not present any qualitative difference; and finally, the lack of dynamic correlation similarly shifts the energies of both S$_1$/S$_0$ MECIs (see Table~S1). Although CASSCF incurs a larger excess energy that shortens excited-state lifetimes~\cite{barbatti2006ultrafast,westermayr2019machine}, it should correctly capture the relative branching ratio between the two decay pathways.

We first performed NAMD simulations using the CASSCF/FSSH method. We generated initial conditions via Wigner sampling at two temperatures, 0~K and 5000~K, simulating 400 and 100 trajectories, respectively. The CASSCF/FSSH trajectories were initiated in the S$_2$ state and propagated with a 0.2 fs time step for the nuclear motion, while the electronic subsystem was integrated with a 100-times smaller time step. From these trajectories, we selected approximately 50,000 configurations to train MACE neural network models for the potential energies and energy gradients of the three lowest singlet states (see SI for details and Fig.~S10). The resulting machine learning (ML) potentials were then employed for extensive LZBL dynamics simulations initiated in both the S$_2$ and S$_1$ states. Additionally, 300 CASSCF/FSSH trajectories with the initial conditions at 0 K were simulated, starting in S$_1$. The rigidity of \mic~ (all vibrational modes are above 900 cm$^{-1}$) rendered its photoinduced dynamics effectively invariant to ground-state temperatures from 0~K to 300~K. The elevated temperature of 5000 K was used to improve the fit quality in the higher energy regions of the configuration space. The ground-state harmonic frequencies and normal modes used for sampling the initial conditions were calculated at the SA(3)-CASSCF(12,12)/aug-cc-pVDZ level of theory and were consistently used throughout the study.

Figure~\ref{fig:jade_dynamics} shows the time-dependent state populations and branching ratios of the resulting photodissociation channels for CASSCF/FSSH dynamics following S$_2$ and S$_1$ photoexcitation at 0~K. Nonadiabatic decay is ultrafast for both excited states. The mean lifetime of the S$_2$ state is 6~fs, with all trajectories transitioning out of S$_2$ within 40~fs. The S$_1$ state exhibits a longer mean lifetime of 18.1 fs when populated from S$_2$ decay, and 21.5~fs when initiated directly. These results are consistent with earlier computational studies~\cite{barbatti2006ultrafast,lan_jade,westermayr2019machine}.

Our dynamics simulations reveal seven distinct photodissociation pathways:
(1) CN bond cleavage: \mic{} $\rightarrow$ CH$_2^+$ $+$ NH$_2$; 
(2) Concerted H$_2$ (2H) elimination from carbon: \mic{} $\rightarrow$ CNH$_2^+$ $+$ H$_2$ (2H); 
(3) Double H-atom elimination from nitrogen: \mic{} $\rightarrow$ NCH$_2^+$ $+$ 2H; 
(4) H-atom elimination from carbon: \mic{} $\rightarrow$ HCNH$_2^+$ $+$ H; 
(5) H-atom elimination from nitrogen: \mic{} $\rightarrow$ H$_2$CNH$^+$ $+$ H; 
(6) Double H-atom elimination from carbon and nitrogen: \mic{} $\rightarrow$ HCNH$^+$ $+$ 2H; 
(7) Non-dissociation: \mic{} remains intact on the simulated timescale. 
The trajectories were assigned to these groups using the procedure outlined in the Supporting Information (see Figs.~S11-S12).

A significantly larger fraction of molecules remains intact after $\sim$100 fs when photoexcitation originates from the S$_1$ state, compared to the S$_2$ state. This difference is attributed to the approximately 1~eV of excess energy available following S$_2$ excitation. Furthermore, the dissociation pathways are rather state-specific: molecules excited to S$_1$ almost exclusively undergo CH bond cleavage, while those excited to S$_2$ predominantly dissociate via CN bond scission or CH bond cleavage. This is in line with the analysis of the potential energy surfaces along the reaction coordinates and the structures of the conical intersections. 

The relative yields of most photodissociation channels depend critically on the total simulation time, as substantial excitation energy drives statistical fragmentation in the hot ground state after internal conversion. In contrast, the direct H$_2$ loss channel, which is mediated by ultrafast decay through a specific MECI, exhibits a constant yield when the simulation time is extended from 50 fs to 100 fs (Fig. S13). This distinct temporal signature provides an effective diagnostic for distinguishing between two mechanisms: direct, MECI-driven decay and indirect, statistical fragmentation.

\begin{table}[t!]
 \begin{threeparttable}
\caption{Fractions of trajectories (\%) decaying via the \spci{} and \ppci{} conical intersections from the CASSCF/FSSH and ML/LZBL simulations. Results are shown for dynamics following S$_1$ and S$_2$ excitation, with and without pre-excitation of the CH$_2$ wagging mode.}
\label{tab:CI_proportions}
\begin{center}

\begin{tabular}{|c|cc|cc|}
\hline
\multirow{2}{*}{}                                          & \multicolumn{2}{c|}{from S$_1$}                              
& \multicolumn{2}{c|}{from S$_2$}                                          \\ \cline{2-5} 
& \multicolumn{1}{c|}{$\sigma\pi^{\ast}$} & \multicolumn{1}{c|}{$\pi\pi^{\ast}$} & \multicolumn{1}{c|}{$\sigma\pi^{\ast}$} & \multicolumn{1}{c|}{$\pi\pi^{\ast}$}  \\ 
\hline
\begin{tabular}[c]{@{}c@{}}CASSCF/FSSH\\ no excitation\end{tabular}                & \multicolumn{1}{c|}{\makecell{ 4.8 \\$\pm$1.2}\tnote{a} }              & \multicolumn{1}{c|}{\makecell{ 74.3 \\$\pm$3.0}\tnote{a} }       & \multicolumn{1}{c|}{1.6}              & \multicolumn{1}{c|}{67.1}   \\ \hline
\begin{tabular}[c]{@{}c@{}}ML/LZBL\\ no excitation\end{tabular}                      & \multicolumn{1}{c|}{3.3}              & \multicolumn{1}{c|}{73.3}   & \multicolumn{1}{c|}{-}                  & \multicolumn{1}{c|}{-}   \\ \hline
\begin{tabular}[c]{@{}c@{}}ML/LZBL\\ double excitation \end{tabular} & \multicolumn{1}{c|}{7.6}              & \multicolumn{1}{c|}{63.2}           & \multicolumn{1}{c|}{-}                  & \multicolumn{1}{c|}{-}           \\
\hline
\end{tabular}

\small
\begin{tablenotes}
\item[a] {Estimated from 5,000 trajectories simulated using the neural network potentials (Section~S10 in the SI) }
\end{tablenotes}
\end{center}
 \end{threeparttable}
\end{table}

\begin{figure*}[t!]
\centering
\includegraphics[width=1.0\textwidth]{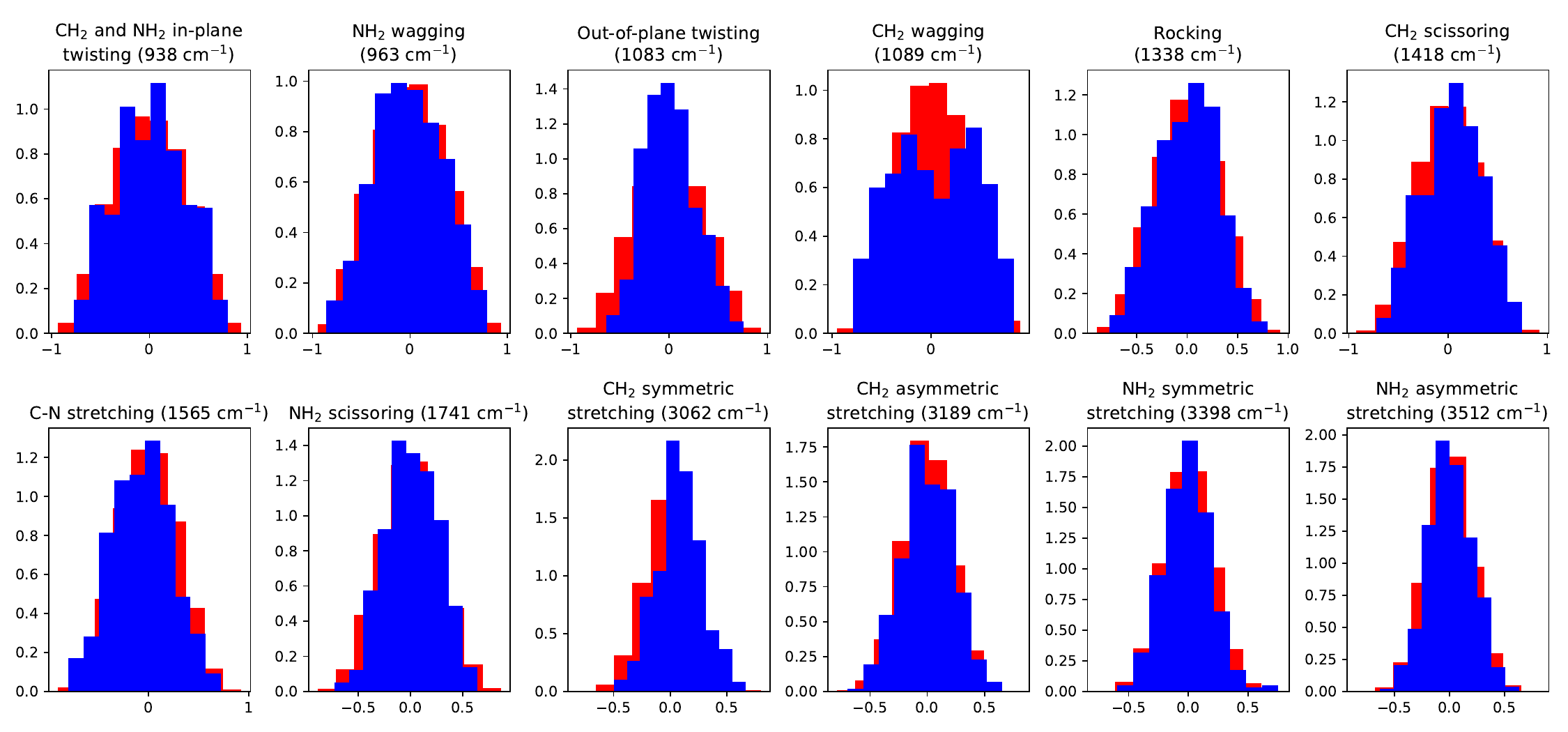}
\caption{Projections of the initial-condition geometries onto the molecular normal modes, color-coded by dynamical outcome. The red distribution represents all 13,000 initial conditions from the ML/LZBL simulations, while the blue distribution corresponds specifically to those trajectories that decay via the \spci{} conical intersection.}
\label{fig:modes_projection}
\end{figure*}

To further explore the S$_1$/S$_0$ decay pathways, we have used our CASSCF-trained neural network potential to locate the nearest MECI for every S$_1$ $\rightarrow$ S$_0$ hop observed in the dynamics (see Section~S10 in the SI). Importantly, the ML potential accurately reproduces MECI geometries and relative energies (Figs.~S14-S15, Table~S2). Table~\ref{tab:CI_proportions} shows the fraction of trajectories from the CASSCF/FSSH NAMD simulations that pass through the \spci{} or \ppci{} conical intersections. The combined branching ratio for the two S$_1$ decay channels is 79.1\% for dynamics initiated in S$_1$, compared to 68.7\% for dynamics initiated in S$_2$. The remaining population of 20.9\% and 31.3\% decays to S$_0$ via regions near the dissociation limits, rather than through the MECIs. 
The higher fraction of dissociation-limited decay for S$_2$ excitation is consistent with the greater initial excess energy available to the system. Remarkably, S$_1$ initiated dynamics exhibit a larger fraction of trajectories decaying through the \spci{} MECI (see Table~\ref{tab:CI_proportions}). This trend aligns well with the distinct sets of Franck-Condon active vibrational modes excited upon transition to S$_1$ and S$_2$ (Fig.~S9).

Given the strong dependence of photochemical quantum yields on initial conditions, we investigate whether selective vibrational excitation can control the branching ratio through the newly discovered \spci{} conical intersection. Although being a minor pathway for both the S$_0$ $\rightarrow$ S$_1$ and S$_0$ $\rightarrow$ S$_2$ excitations, a larger fraction of trajectories decays through the \spci{} MECI when the dynamics is initiated in S$_1$ (4.8 $\pm$ 1.2 \%). This enables to achieve a statistical confidence in the calculated branching ratios (Table~\ref{tab:CI_proportions}). Furthermore, vibrational pre-excitation deposits additional excess energy. When combined with the inherently larger excess energy of the S$_0$ $\rightarrow$ S$_2$ excitation compared to S$_0$ $\rightarrow$ S$_1$, this results in a higher propensity for non-selective photoinduced dissociation, even though ultrafast internal conversion is expected to occur on a timescale competitive with intramolecular vibrational redistribution. Our analysis of vibrational pre-excitation is therefore confined to dynamics initiated in the S$_1$ state.

Approximately 13,000 initial conditions were sampled from the Wigner distribution around the S$_0$ equilibrium geometry, a number sufficient to achieve statistical convergence for the conical intersection branching ratios. Based on the ML/LZBL NAMD simulations starting in S$_1$, 13,000 trajectories are then classified as passing through either the \spci{} or \ppci{} conical intersection using the aforementioned procedure. When benchmarked against CASSCF/FSSH, the ML/LZBL dynamics show excellent agreement, accurately reproducing both the branching ratio (Table~\ref{tab:CI_proportions}) and, when corrected via out-of-sample diagnostics, the excited-state lifetimes (Section~S12 in the SI, Figs.~S16-S18).

\begin{figure*}[t!]
\centering
\includegraphics[width=0.8\textwidth]{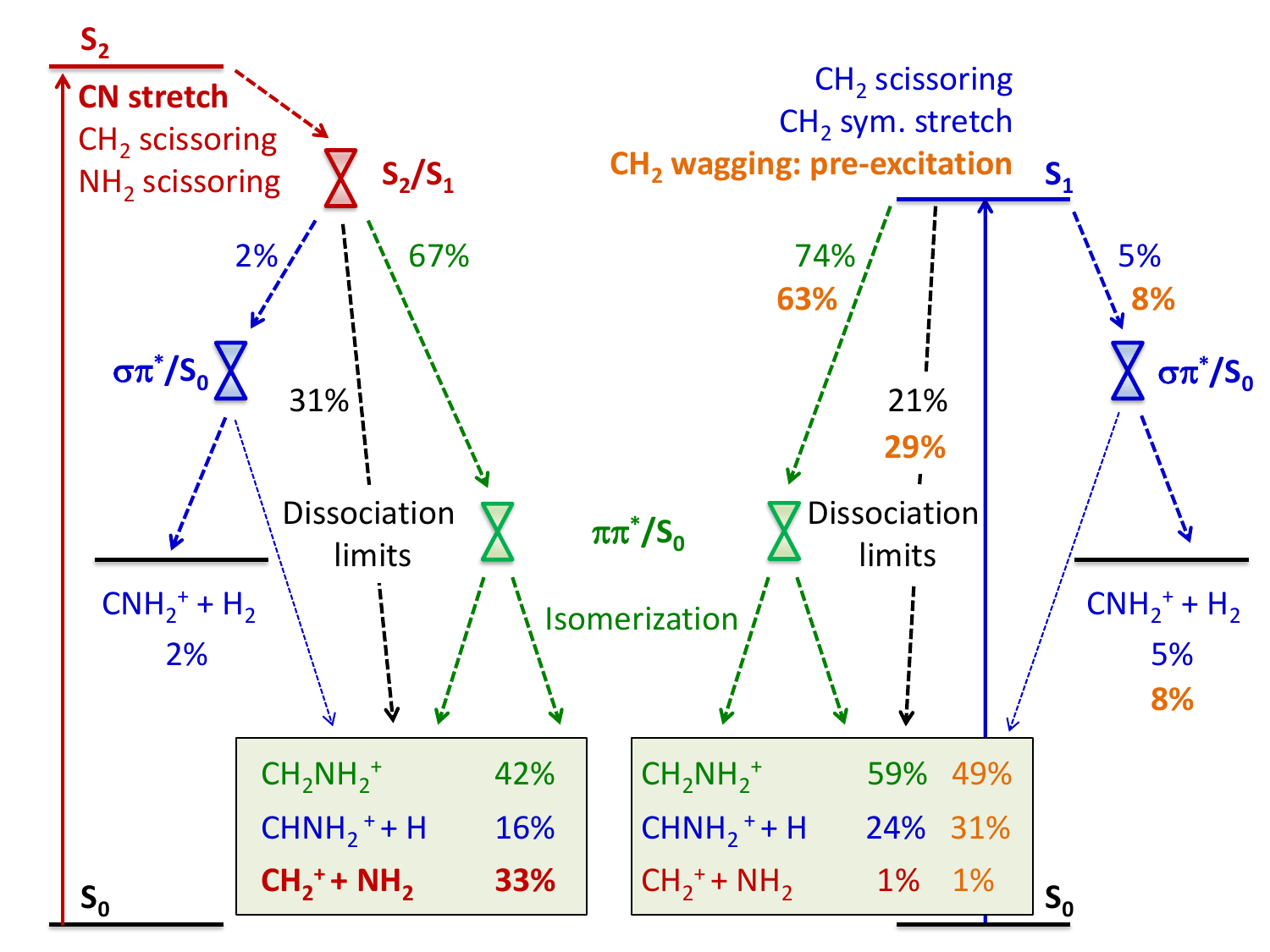}
\caption{Schematic overview of the major excited-state decay pathways in \mic{}, showing branching ratios and the key conical intersections that mediate the nonadiabatic dynamics. Branching ratios are determined from CASSCF/FSSH dynamics (analyzing hops via the nearest MECI) and ML/LZBL dynamics (analyzing products formed within 100 fs). The Franck-Condon active modes and the pre-excited mode are indicated.}
\label{fig:fscheme}
\end{figure*}

Figure~\ref{fig:modes_projection} compares the projections of all initial conditions onto the molecular normal modes (red) against the subset that channels dynamics through the \spci{} MECI (blue). The pre-excitation of the CH$_2$ wagging mode (1089 cm$^{-1}$) creates a pronounced bias toward the \spci{} conical intersection. This is evidenced by a bimodal distribution in the initial conditions that channels reactive trajectories away from the equilibrium geometry. The efficacy of this mode stems from its direct connection to CH$_2$ pyramidalization, which is a key coordinate for reaching the \spci{} MECI. Furthermore, while this mode is Franck-Condon inactive upon the S$_0$ $\rightarrow$ S$_1$ transition, its selective pre-excitation performs a dual function: it not only provides the necessary vibrational energy but also vibronically activates the otherwise forbidden S$_0$ $\rightarrow$ S$_1$ excitation (a reduction in molecular symmetry from C$_{2v}$ to C$_s$), creating a directed pathway to the target intersection. In contrast, the out-of-plane twisting mode (1083 cm$^{-1}$) associated with torsion around the CN bond shows a narrower distribution for the subset leading to the \spci{} MECI than for the total set of initial conditions. Pre-excitation of this mode, instead, directs the dynamics through the main branch via the \ppci{} MECI. These two modes with nearly degenerate frequencies guide the dynamics into distinct relaxation pathways: the IR-active B$_1$ (CH$_2$ wagging) mode to the \spci{} channel and the IR-inactive A$_2$ (out-of-plane twisting) mode to the \ppci{} channel. Due to their different symmetries, the CH$_2$ wagging mode can be pumped  independently by direct IR excitation (Fig.~S20).

To model the dynamics with selective pre-excitation, initial conditions were sampled from the Wigner distribution for a doubly excited CH$_2$ wagging vibrational mode, and the results are compared to the non-pre-excited case. Both datasets comprise 13,000 trajectories. Table \ref{tab:CI_proportions} summarizes how pre-excitation of the CH$_2$ wagging mode affects the branching ratios for the \spci{} and \ppci{} decay channels in ML/LZBL dynamics.  Importantly, pre-excitation acts as a control mechanism, suppressing the \ppci{} channel by 10\% and enhancing the \spci{} channel by a factor of two.

Figure~\ref{fig:fscheme} summarizes the major excited-state decay channels and their branching ratios, determined by analyzing hops to the ground state via the nearest MECI or the final products (see also Fig.~S19). The novel \spci{} conical intersection exclusively leads to photodissociation with direct H$_2$ loss. In contrast, other dissociation channels occur either in the hot ground state after passing through the \ppci{} conical intersection or near the dissociation limits. Furthermore, the initial excitation dictates the primary photodissociation pathway: CN bond cleavage is promoted by S$_2$ excitation, whereas single H-atom loss is enhanced by S$_1$ excitation. This selectivity is governed by their distinct Franck-Condon active modes. The vibronically allowed transition, mediated by the CH$_2$ wagging mode, populates the S$_1$ state in \mic{}. The ML-accelerated nonadiabatic dynamics simulations reveal that selective excitation of this mode funnels the excited-state population through the newly discovered \spci{} conical intersection, directly leading to H$_2$ loss and yielding carbene CNH$_2^+$ as a direct precursor to HCNH$^+$. Trajectory analysis confirms this new photochemical pathway, with an illustrative animation provided in the SI.

In summary, we report a new photochemical pathway in \mic{} -- direct H$_2$ elimination mediated by a \spci{} conical intersection, significantly advancing the understanding of this fundamental model system. This discovery is enabled by a novel methodology that successfully combines high-level \textit{ab initio} theory with machine-learning accelerated dynamics, demonstrating that accurate electronic structure is indispensable for reliably capturing such complex nonadiabatic events. We have quantified this decay channel and established a clear precedent for mode-specific vibrational control by showing that vibrational pre-excitation selectively funnels the dynamics toward this predetermined outcome. The ultrafast, non-statistical nature of this reaction distinguishes it fundamentally from conventional statistical dissociation processes. The implications of this newly discovered pathway extend from a fundamental model system to planetary science. For Titan's atmosphere, where \mic{} is a known precursor to HCNH$^+$, this mechanism provides a photochemical route to H$_2$ elimination that could directly influence the ionospheric reaction network. More broadly, the general methodology developed here for mapping trajectories to conical intersections provides a powerful and transferable approach for unraveling and controlling complex photochemistry across diverse molecular systems.

\begin{acknowledgement}

This work was supported by the Russian Science Foundation (Grant No. 24-43-00041). The calculations were carried out using the equipment of the shared research facilities of HPC computing resources at Lomonosov Moscow State University as well as the local resources (RSC Tornado) provided through the Lomonosov Moscow State University Program of Development.
\end{acknowledgement}

\begin{suppinfo}

Vertical transitions, structures of MECIs, topographies around MECIs, potential energy surface scans,  Franck-Condon active vibrational modes, details of nonadiabatic dynamics simulations, dataset construction, neural network potentials, analysis of photodissociation channels and their temporal evolution, ML-based analysis of excited-state decay through MECIs, comparison of \textit{ab initio} and ML population dynamics, out-of-sample diagnostics, IR spectrum, animations of the branching plane coordinates and a representative nonadiabatic trajectory through the newly discovered \spci{} conical intersection.

\end{suppinfo}


\begin{mcitethebibliography}{60}
\providecommand*\natexlab[1]{#1}
\providecommand*\mciteSetBstSublistMode[1]{}
\providecommand*\mciteSetBstMaxWidthForm[2]{}
\providecommand*\mciteBstWouldAddEndPuncttrue
  {\def\EndOfBibitem{\unskip.}}
\providecommand*\mciteBstWouldAddEndPunctfalse
  {\let\EndOfBibitem\relax}
\providecommand*\mciteSetBstMidEndSepPunct[3]{}
\providecommand*\mciteSetBstSublistLabelBeginEnd[3]{}
\providecommand*\EndOfBibitem{}
\mciteSetBstSublistMode{f}
\mciteSetBstMaxWidthForm{subitem}{(\alph{mcitesubitemcount})}
\mciteSetBstSublistLabelBeginEnd
  {\mcitemaxwidthsubitemform\space}
  {\relax}
  {\relax}

\bibitem[Polli \latin{et~al.}(2010)Polli, Alto{\`e}, Weingart, Spillane,
  Manzoni, Brida, Tomasello, Orlandi, Kukura, Mathies, \latin{et~al.}
  others]{polli2010conical}
Polli,~D.; Alto{\`e},~P.; Weingart,~O.; Spillane,~K.~M.; Manzoni,~C.;
  Brida,~D.; Tomasello,~G.; Orlandi,~G.; Kukura,~P.; Mathies,~R.~A.
  \latin{et~al.}  Conical intersection dynamics of the primary
  photoisomerization event in vision. \emph{Nature} \textbf{2010}, \emph{467},
  440--443\relax
\mciteBstWouldAddEndPuncttrue
\mciteSetBstMidEndSepPunct{\mcitedefaultmidpunct}
{\mcitedefaultendpunct}{\mcitedefaultseppunct}\relax
\EndOfBibitem
\bibitem[Warshel(1976)]{warshel1976bicycle}
Warshel,~A. Bicycle-pedal model for the first step in the vision process.
  \emph{Nature} \textbf{1976}, \emph{260}, 679--683\relax
\mciteBstWouldAddEndPuncttrue
\mciteSetBstMidEndSepPunct{\mcitedefaultmidpunct}
{\mcitedefaultendpunct}{\mcitedefaultseppunct}\relax
\EndOfBibitem
\bibitem[Karkas \latin{et~al.}(2016)Karkas, Porco~Jr, and
  Stephenson]{karkas2016photochemical}
Karkas,~M.~D.; Porco~Jr,~J.~A.; Stephenson,~C.~R. Photochemical approaches to
  complex chemotypes: applications in natural product synthesis. \emph{Chem.
  Rev.} \textbf{2016}, \emph{116}, 9683--9747\relax
\mciteBstWouldAddEndPuncttrue
\mciteSetBstMidEndSepPunct{\mcitedefaultmidpunct}
{\mcitedefaultendpunct}{\mcitedefaultseppunct}\relax
\EndOfBibitem
\bibitem[Hoffmann(2008)]{hoffmann2008photochemical}
Hoffmann,~N. Photochemical reactions as key steps in organic synthesis.
  \emph{Chem. Rev.} \textbf{2008}, \emph{108}, 1052--1103\relax
\mciteBstWouldAddEndPuncttrue
\mciteSetBstMidEndSepPunct{\mcitedefaultmidpunct}
{\mcitedefaultendpunct}{\mcitedefaultseppunct}\relax
\EndOfBibitem
\bibitem[Long \latin{et~al.}(2017)Long, Prezhdo, and
  Fang]{long2017nonadiabatic}
Long,~R.; Prezhdo,~O.~V.; Fang,~W. Nonadiabatic charge dynamics in novel solar
  cell materials. \emph{Wiley Interdiscip. Rev.:Comput. Mol. Sci.}
  \textbf{2017}, \emph{7}, e1305\relax
\mciteBstWouldAddEndPuncttrue
\mciteSetBstMidEndSepPunct{\mcitedefaultmidpunct}
{\mcitedefaultendpunct}{\mcitedefaultseppunct}\relax
\EndOfBibitem
\bibitem[Zhang \latin{et~al.}(2013)Zhang, Zou, and Tian]{zhang2013photochromic}
Zhang,~J.; Zou,~Q.; Tian,~H. Photochromic materials: more than meets the eye.
  \emph{Adv. Mater.} \textbf{2013}, \emph{25}, 378--399\relax
\mciteBstWouldAddEndPuncttrue
\mciteSetBstMidEndSepPunct{\mcitedefaultmidpunct}
{\mcitedefaultendpunct}{\mcitedefaultseppunct}\relax
\EndOfBibitem
\bibitem[Liu \latin{et~al.}(1990)Liu, Hashimoto, and
  Fujishima]{liu1990photoelectrochemical}
Liu,~Z.~F.; Hashimoto,~K.; Fujishima,~A. Photoelectrochemical information
  storage using an azobenzene derivative. \emph{Nature} \textbf{1990},
  \emph{347}, 658--660\relax
\mciteBstWouldAddEndPuncttrue
\mciteSetBstMidEndSepPunct{\mcitedefaultmidpunct}
{\mcitedefaultendpunct}{\mcitedefaultseppunct}\relax
\EndOfBibitem
\bibitem[Curchod and Martínez(2018)Curchod, and
  Martínez]{Martinez2018_review}
Curchod,~B. F.~E.; Martínez,~T.~J. Ab Initio Nonadiabatic Quantum Molecular
  Dynamics. \emph{Chem. Rev.} \textbf{2018}, \emph{118}, 3305--3336\relax
\mciteBstWouldAddEndPuncttrue
\mciteSetBstMidEndSepPunct{\mcitedefaultmidpunct}
{\mcitedefaultendpunct}{\mcitedefaultseppunct}\relax
\EndOfBibitem
\bibitem[Meyer and Worth(2003)Meyer, and Worth]{Worth2003}
Meyer,~H.-D.; Worth,~G.~A. Quantum molecular dynamics: propagating wavepackets
  and density operators using the multiconfiguration time-dependent Hartree
  method. \emph{Theor. Chem. Acc.} \textbf{2003}, \emph{109}, 251--267\relax
\mciteBstWouldAddEndPuncttrue
\mciteSetBstMidEndSepPunct{\mcitedefaultmidpunct}
{\mcitedefaultendpunct}{\mcitedefaultseppunct}\relax
\EndOfBibitem
\bibitem[Martinez \latin{et~al.}(1996)Martinez, Ben-Nun, and
  Levine]{Martinez1996}
Martinez,~T.~J.; Ben-Nun,~M.; Levine,~R.~D. Multi-Electronic-State Molecular
  Dynamics: A Wave Function Approach with Applications. \emph{J. Phys. Chem.} \textbf{1996}, \emph{100}, 7884--7895\relax
\mciteBstWouldAddEndPuncttrue
\mciteSetBstMidEndSepPunct{\mcitedefaultmidpunct}
{\mcitedefaultendpunct}{\mcitedefaultseppunct}\relax
\EndOfBibitem
\bibitem[Ben-Nun \latin{et~al.}(2000)Ben-Nun, Quenneville, and
  Martínez]{Martinez2000}
Ben-Nun,~M.; Quenneville,~J.; Martínez,~T.~J. Ab Initio Multiple Spawning:
  Photochemistry from First Principles Quantum Molecular Dynamics. \emph{
  J. Phys. Chem. A} \textbf{2000}, \emph{104}, 5161--5175\relax
\mciteBstWouldAddEndPuncttrue
\mciteSetBstMidEndSepPunct{\mcitedefaultmidpunct}
{\mcitedefaultendpunct}{\mcitedefaultseppunct}\relax
\EndOfBibitem
\bibitem[C.~Tully(1998)]{Tully1998}
C.~Tully,~J. Mixed quantum–classical dynamics. \emph{Faraday Discuss.}
  \textbf{1998}, \emph{110}, 407--419\relax
\mciteBstWouldAddEndPuncttrue
\mciteSetBstMidEndSepPunct{\mcitedefaultmidpunct}
{\mcitedefaultendpunct}{\mcitedefaultseppunct}\relax
\EndOfBibitem
\bibitem[Crespo-Otero and Barbatti(2018)Crespo-Otero, and
  Barbatti]{Barbatti2018}
Crespo-Otero,~R.; Barbatti,~M. Recent advances and perspectives on nonadiabatic
  mixed quantum–classical dynamics. \emph{Chem. Rev.} \textbf{2018},
  \emph{118}, 7026--7068\relax
\mciteBstWouldAddEndPuncttrue
\mciteSetBstMidEndSepPunct{\mcitedefaultmidpunct}
{\mcitedefaultendpunct}{\mcitedefaultseppunct}\relax
\EndOfBibitem
\bibitem[Subotnik \latin{et~al.}(2016)Subotnik, Jain, Landry, Petit, Ouyang,
  and Bellonzi]{Subotnik2016}
Subotnik,~J.~E.; Jain,~A.; Landry,~B.; Petit,~A.; Ouyang,~W.; Bellonzi,~N.
  Understanding the Surface Hopping View of Electronic Transitions and
  Decoherence. \emph{Ann. Rev. Phys. Chem.} \textbf{2016},
  \emph{67}, 387--417\relax
\mciteBstWouldAddEndPuncttrue
\mciteSetBstMidEndSepPunct{\mcitedefaultmidpunct}
{\mcitedefaultendpunct}{\mcitedefaultseppunct}\relax
\EndOfBibitem
\bibitem[Vindel-Zandbergen \latin{et~al.}(2021)Vindel-Zandbergen, Ibele, Ha,
  Min, Curchod, and Maitra]{AIMS_TSH_D}
Vindel-Zandbergen,~P.; Ibele,~L.~M.; Ha,~J.-K.; Min,~S.~K.; Curchod,~B. F.~E.;
  Maitra,~N.~T. Study of the Decoherence Correction Derived from the Exact
  Factorization Approach for Nonadiabatic Dynamics. \emph{J. Chem.
  Theo. Comput.} \textbf{2021}, \emph{17}, 3852--3862\relax
\mciteBstWouldAddEndPuncttrue
\mciteSetBstMidEndSepPunct{\mcitedefaultmidpunct}
{\mcitedefaultendpunct}{\mcitedefaultseppunct}\relax
\EndOfBibitem
\bibitem[Gozem \latin{et~al.}(2017)Gozem, Luk, Schapiro, and
  Olivucci]{Gozem2017}
Gozem,~S.; Luk,~H.~L.; Schapiro,~I.; Olivucci,~M. Theory and simulation of the
  ultrafast double-bond isomerization of biological chromophores. \emph{Chem.
  Rev.} \textbf{2017}, \emph{117}, 13502--13565\relax
\mciteBstWouldAddEndPuncttrue
\mciteSetBstMidEndSepPunct{\mcitedefaultmidpunct}
{\mcitedefaultendpunct}{\mcitedefaultseppunct}\relax
\EndOfBibitem
\bibitem[Bochenkova(2024)]{BOCHENKOVA2024141}
Bochenkova,~A.~V. In \emph{Comprehensive Computational Chemistry (First
  Edition)}, first edition ed.; Yáñez,~M., Boyd,~R.~J., Eds.; Elsevier:
  Oxford, 2024; pp 141--157\relax
\mciteBstWouldAddEndPuncttrue
\mciteSetBstMidEndSepPunct{\mcitedefaultmidpunct}
{\mcitedefaultendpunct}{\mcitedefaultseppunct}\relax
\EndOfBibitem
\bibitem[Bar and Rosenwaks(2001)Bar, and Rosenwaks]{Bar2001}
Bar,~I.; Rosenwaks,~S. Controlling bond cleavage and probing intramolecular
  dynamics via photodissociation of rovibrationally excited molecules.
  \emph{Int. Rev. Phys. Chem.} \textbf{2001}, \emph{20}, 711--749\relax
\mciteBstWouldAddEndPuncttrue
\mciteSetBstMidEndSepPunct{\mcitedefaultmidpunct}
{\mcitedefaultendpunct}{\mcitedefaultseppunct}\relax
\EndOfBibitem
\bibitem[Robertson and Worth(2017)Robertson, and Worth]{Worth2017}
Robertson,~C.; Worth,~G.~A. Modelling the vibrationally mediated
  photo-dissociation of acetylene. \emph{Phys. Chem. Chem. Phys.}
  \textbf{2017}, \emph{19}, 29483--29497\relax
\mciteBstWouldAddEndPuncttrue
\mciteSetBstMidEndSepPunct{\mcitedefaultmidpunct}
{\mcitedefaultendpunct}{\mcitedefaultseppunct}\relax
\EndOfBibitem
\bibitem[Dral \latin{et~al.}(2018)Dral, Barbatti, and Thiel]{Dral2018}
Dral,~P.~O.; Barbatti,~M.; Thiel,~W. Nonadiabatic Excited-State Dynamics with
  Machine Learning. \emph{J. Phys. Chem. Lett.} \textbf{2018}, \emph{9},
  5660--5663\relax
\mciteBstWouldAddEndPuncttrue
\mciteSetBstMidEndSepPunct{\mcitedefaultmidpunct}
{\mcitedefaultendpunct}{\mcitedefaultseppunct}\relax
\EndOfBibitem
\bibitem[Hu \latin{et~al.}(2018)Hu, Xie, Li, Li, and Lan]{Lan2018}
Hu,~D.; Xie,~Y.; Li,~X.; Li,~L.; Lan,~Z. Inclusion of Machine Learning Kernel
  Ridge Regression Potential Energy Surfaces in On-the-Fly Nonadiabatic
  Molecular Dynamics Simulation. \emph{J. Phys. Chem. Lett.} \textbf{2018},
  \emph{9}, 2725--2732\relax
\mciteBstWouldAddEndPuncttrue
\mciteSetBstMidEndSepPunct{\mcitedefaultmidpunct}
{\mcitedefaultendpunct}{\mcitedefaultseppunct}\relax
\EndOfBibitem
\bibitem[Westermayr and Marquetand(2021)Westermayr, and
  Marquetand]{Marquetand2021}
Westermayr,~J.; Marquetand,~P. Machine Learning for Electronically Excited
  States of Molecules. \emph{Chem. Rev.} \textbf{2021}, \emph{121},
  9873--9926\relax
\mciteBstWouldAddEndPuncttrue
\mciteSetBstMidEndSepPunct{\mcitedefaultmidpunct}
{\mcitedefaultendpunct}{\mcitedefaultseppunct}\relax
\EndOfBibitem
\bibitem[Hou \latin{et~al.}(2024)Hou, Zhang, Zhang, Ge, and Dral]{Dral2024}
Hou,~Y.-F.; Zhang,~L.; Zhang,~Q.; Ge,~F.; Dral,~P.~O. Physics-informed active
  learning for accelerating quantum chemical simulations. \emph{J. Chem. Theo.
  Comput.} \textbf{2024}, \emph{20}, 7744--7754\relax
\mciteBstWouldAddEndPuncttrue
\mciteSetBstMidEndSepPunct{\mcitedefaultmidpunct}
{\mcitedefaultendpunct}{\mcitedefaultseppunct}\relax
\EndOfBibitem
\bibitem[Sr\v{s}e\v{n} \latin{et~al.}(2024)Sr\v{s}e\v{n}, von Lilienfeld, and
  Slav\'{i}\v{c}ek]{Slavicek2024}
Sr\v{s}e\v{n},~{\v{S}}.; von Lilienfeld,~O.~A.; Slav\'{i}\v{c}ek,~P. Fast and
  accurate excited states predictions: machine learning and diabatization.
  \emph{Phys. Chem. Chem. Phys.} \textbf{2024}, \emph{26}, 4306--4319\relax
\mciteBstWouldAddEndPuncttrue
\mciteSetBstMidEndSepPunct{\mcitedefaultmidpunct}
{\mcitedefaultendpunct}{\mcitedefaultseppunct}\relax
\EndOfBibitem
\bibitem[Kocer \latin{et~al.}(2022)Kocer, Ko, and Behler]{Jorg2022}
Kocer,~E.; Ko,~T.~W.; Behler,~J. Neural network potentials: a concise overview
  of methods. \emph{Annu. Rev. Phys. Chem.} \textbf{2022}, \emph{73},
  163--186\relax
\mciteBstWouldAddEndPuncttrue
\mciteSetBstMidEndSepPunct{\mcitedefaultmidpunct}
{\mcitedefaultendpunct}{\mcitedefaultseppunct}\relax
\EndOfBibitem
\bibitem[Hoon~Choi \latin{et~al.}(2001)Hoon~Choi, Tae~Park, and
  Soo~Kim]{hoon2001theoretical}
Hoon~Choi,~T.; Tae~Park,~S.; Soo~Kim,~M. Theoretical and experimental studies
  of the dissociation dynamics of methaniminium cation, CH$_2$NH$_2^+
  \leftarrow$ CHNH$^+$+ H$_2$: reaction path bifurcation. \emph{J. Chem. Phys.}
  \textbf{2001}, \emph{114}, 6051--6057\relax
\mciteBstWouldAddEndPuncttrue
\mciteSetBstMidEndSepPunct{\mcitedefaultmidpunct}
{\mcitedefaultendpunct}{\mcitedefaultseppunct}\relax
\EndOfBibitem
\bibitem[Suarez and Sordo(1997)Suarez, and Sordo]{suarez1997ab}
Suarez,~D.; Sordo,~T. Ab initio study of the H2 elimination from CH$_2$OH$^+$,
  CH$_2$NH$_2^+$, and CH$_2$SH$^+$. \emph{J. Phys. Chem. A} \textbf{1997},
  \emph{101}, 1561--1566\relax
\mciteBstWouldAddEndPuncttrue
\mciteSetBstMidEndSepPunct{\mcitedefaultmidpunct}
{\mcitedefaultendpunct}{\mcitedefaultseppunct}\relax
\EndOfBibitem
\bibitem[Barbatti \latin{et~al.}(2007)Barbatti, Granucci, Persico, Ruckenbauer,
  Vazdar, Eckert-Maksi{\'c}, and Lischka]{barbatti2007fly}
Barbatti,~M.; Granucci,~G.; Persico,~M.; Ruckenbauer,~M.; Vazdar,~M.;
  Eckert-Maksi{\'c},~M.; Lischka,~H. The on-the-fly surface-hopping program
  system Newton-X: Application to ab initio simulation of the nonadiabatic
  photodynamics of benchmark systems. \emph{J. Photochem. Photobiol. A}
  \textbf{2007}, \emph{190}, 228--240\relax
\mciteBstWouldAddEndPuncttrue
\mciteSetBstMidEndSepPunct{\mcitedefaultmidpunct}
{\mcitedefaultendpunct}{\mcitedefaultseppunct}\relax
\EndOfBibitem
\bibitem[Tapavicza \latin{et~al.}(2007)Tapavicza, Tavernelli, and
  Rothlisberger]{tapavicza2007trajectory}
Tapavicza,~E.; Tavernelli,~I.; Rothlisberger,~U. Trajectory Surface Hopping
  within Linear Response Time-Dependent Density-Functional Theory. \emph{Phys.
  Rev. Lett.} \textbf{2007}, \emph{98}, 023001\relax
\mciteBstWouldAddEndPuncttrue
\mciteSetBstMidEndSepPunct{\mcitedefaultmidpunct}
{\mcitedefaultendpunct}{\mcitedefaultseppunct}\relax
\EndOfBibitem
\bibitem[Yamazaki and Kato(2005)Yamazaki, and Kato]{yamazaki2005locating}
Yamazaki,~S.; Kato,~S. Locating the lowest free-energy point on conical
  intersection in polar solvent: Reference interaction site model
  self-consistent field study of ethylene and CH$_2$NH$_2^+$. \emph{J. Chem.
  Phys.} \textbf{2005}, \emph{123}, 114510\relax
\mciteBstWouldAddEndPuncttrue
\mciteSetBstMidEndSepPunct{\mcitedefaultmidpunct}
{\mcitedefaultendpunct}{\mcitedefaultseppunct}\relax
\EndOfBibitem
\bibitem[Suchan \latin{et~al.}(2020)Suchan, Jano\v{s}, and
  Slav\'{i}\v{c}ek]{suchan2020pragmatic}
Suchan,~J.; Jano\v{s},~J.; Slav\'{i}\v{c}ek,~P. Pragmatic approach to
  photodynamics: Mixed Landau--Zener surface hopping with intersystem crossing.
  \emph{J. Chem. Theo. Comput.} \textbf{2020}, \emph{16}, 5809--5820\relax
\mciteBstWouldAddEndPuncttrue
\mciteSetBstMidEndSepPunct{\mcitedefaultmidpunct}
{\mcitedefaultendpunct}{\mcitedefaultseppunct}\relax
\EndOfBibitem
\bibitem[Pittner \latin{et~al.}(2009)Pittner, Lischka, and
  Barbatti]{pittner2009optimization}
Pittner,~J.; Lischka,~H.; Barbatti,~M. Optimization of mixed quantum-classical
  dynamics: Time-derivative coupling terms and selected couplings. \emph{Chem.
  Phys.} \textbf{2009}, \emph{356}, 147--152\relax
\mciteBstWouldAddEndPuncttrue
\mciteSetBstMidEndSepPunct{\mcitedefaultmidpunct}
{\mcitedefaultendpunct}{\mcitedefaultseppunct}\relax
\EndOfBibitem
\bibitem[West \latin{et~al.}(2014)West, Barbatti, Lischka, and
  Windus]{west2014nonadiabatic}
West,~A.~C.; Barbatti,~M.; Lischka,~H.; Windus,~T.~L. Nonadiabatic dynamics
  study of methaniminium with ORMAS: Challenges of incomplete active spaces in
  dynamics simulations. \emph{Comput. Theor. Chem.} \textbf{2014}, \emph{1040},
  158--166\relax
\mciteBstWouldAddEndPuncttrue
\mciteSetBstMidEndSepPunct{\mcitedefaultmidpunct}
{\mcitedefaultendpunct}{\mcitedefaultseppunct}\relax
\EndOfBibitem
\bibitem[Barbatti \latin{et~al.}(2006)Barbatti, Aquino, and
  Lischka]{barbatti2006ultrafast}
Barbatti,~M.; Aquino,~A.~J.; Lischka,~H. Ultrafast two-step process in the
  non-adiabatic relaxation of the CH$_2$NH$_2$ molecule. \emph{Mol. Phys.}
  \textbf{2006}, \emph{104}, 1053--1060\relax
\mciteBstWouldAddEndPuncttrue
\mciteSetBstMidEndSepPunct{\mcitedefaultmidpunct}
{\mcitedefaultendpunct}{\mcitedefaultseppunct}\relax
\EndOfBibitem
\bibitem[Fabiano \latin{et~al.}(2008)Fabiano, Groenhof, and
  Thiel]{fabiano2008approximate}
Fabiano,~E.; Groenhof,~G.; Thiel,~W. Approximate switching algorithms for
  trajectory surface hopping. \emph{Chem. Phys.} \textbf{2008}, \emph{351},
  111--116\relax
\mciteBstWouldAddEndPuncttrue
\mciteSetBstMidEndSepPunct{\mcitedefaultmidpunct}
{\mcitedefaultendpunct}{\mcitedefaultseppunct}\relax
\EndOfBibitem
\bibitem[Westermayr \latin{et~al.}(2019)Westermayr, Gastegger, Menger, Mai,
  Gonz{\'a}lez, and Marquetand]{westermayr2019machine}
Westermayr,~J.; Gastegger,~M.; Menger,~M.~F.; Mai,~S.; Gonz{\'a}lez,~L.;
  Marquetand,~P. Machine learning enables long time scale molecular
  photodynamics simulations. \emph{Chem. Sci.} \textbf{2019}, \emph{10},
  8100--8107\relax
\mciteBstWouldAddEndPuncttrue
\mciteSetBstMidEndSepPunct{\mcitedefaultmidpunct}
{\mcitedefaultendpunct}{\mcitedefaultseppunct}\relax
\EndOfBibitem
\bibitem[Westermayr \latin{et~al.}(2020)Westermayr, Gastegger, and
  Marquetand]{westermayr2020combining}
Westermayr,~J.; Gastegger,~M.; Marquetand,~P. Combining SchNet and SHARC: The
  SchNarc machine learning approach for excited-state dynamics. \emph{J. Phys.
  Chem. Lett.} \textbf{2020}, \emph{11}, 3828--3834\relax
\mciteBstWouldAddEndPuncttrue
\mciteSetBstMidEndSepPunct{\mcitedefaultmidpunct}
{\mcitedefaultendpunct}{\mcitedefaultseppunct}\relax
\EndOfBibitem
\bibitem[Du and Lan(2015)Du, and Lan]{lan_jade}
Du,~L.; Lan,~Z. An On-the-Fly Surface-Hopping Program JADE for nonadiabatic
  molecular dynamics of polyatomic systems: implementation and applications.
  \emph{J. Chem. Theo. Comput.} \textbf{2015}, \emph{11}, 1360--1374, PMID:
  26574348\relax
\mciteBstWouldAddEndPuncttrue
\mciteSetBstMidEndSepPunct{\mcitedefaultmidpunct}
{\mcitedefaultendpunct}{\mcitedefaultseppunct}\relax
\EndOfBibitem
\bibitem[Barbatti \latin{et~al.}(2008)Barbatti, Ruckenbauer, Szymczak, Aquino,
  and Lischka]{barbatti2008nonadiabatic}
Barbatti,~M.; Ruckenbauer,~M.; Szymczak,~J.~J.; Aquino,~A.~J.; Lischka,~H.
  Nonadiabatic excited-state dynamics of polar $\pi$-systems and related model
  compounds of biological relevance. \emph{Phys. Chem. Chem. Phys.}
  \textbf{2008}, \emph{10}, 482--494\relax
\mciteBstWouldAddEndPuncttrue
\mciteSetBstMidEndSepPunct{\mcitedefaultmidpunct}
{\mcitedefaultendpunct}{\mcitedefaultseppunct}\relax
\EndOfBibitem
\bibitem[Nixon(2024)]{Nixon2024}
Nixon,~C.~A. The composition and chemistry of Titan’s atmosphere. \emph{ACS
  Earth Space Chem.} \textbf{2024}, \emph{8}, 406--456\relax
\mciteBstWouldAddEndPuncttrue
\mciteSetBstMidEndSepPunct{\mcitedefaultmidpunct}
{\mcitedefaultendpunct}{\mcitedefaultseppunct}\relax
\EndOfBibitem
\bibitem[Singh \latin{et~al.}(2010)Singh, Shen, Zhou, Schlegel, and
  Suits]{singh2010photodissociation}
Singh,~P.~C.; Shen,~L.; Zhou,~J.; Schlegel,~H.~B.; Suits,~A.~G.
  Photodissociation dynamics of methylamine cation and its relevance to Titan's
  ionosphere. \emph{Astrophys. J.} \textbf{2010}, \emph{710}, 112\relax
\mciteBstWouldAddEndPuncttrue
\mciteSetBstMidEndSepPunct{\mcitedefaultmidpunct}
{\mcitedefaultendpunct}{\mcitedefaultseppunct}\relax
\EndOfBibitem
\bibitem[Pei and Farrar(2012)Pei, and Farrar]{pei2012ion}
Pei,~L.; Farrar,~J.~M. Ion imaging study of reaction dynamics in the N$^+$+
  CH$_4$ system. \emph{J. Chem. Phys.} \textbf{2012}, \emph{137}, 154312\relax
\mciteBstWouldAddEndPuncttrue
\mciteSetBstMidEndSepPunct{\mcitedefaultmidpunct}
{\mcitedefaultendpunct}{\mcitedefaultseppunct}\relax
\EndOfBibitem
\bibitem[Thackston and Fortenberry(2018)Thackston, and
  Fortenberry]{thackston2018quantum}
Thackston,~R.; Fortenberry,~R.~C. Quantum chemical spectral characterization of
  {CH$_2$NH$_2^+$} for remote sensing of Titan’s atmosphere. \emph{Icarus}
  \textbf{2018}, \emph{299}, 187--193\relax
\mciteBstWouldAddEndPuncttrue
\mciteSetBstMidEndSepPunct{\mcitedefaultmidpunct}
{\mcitedefaultendpunct}{\mcitedefaultseppunct}\relax
\EndOfBibitem
\bibitem[Vuitton \latin{et~al.}(2006)Vuitton, Yelle, and
  Anicich]{vuitton2006nitrogen}
Vuitton,~V.; Yelle,~R.; Anicich,~V. The nitrogen chemistry of Titan’s upper
  atmosphere revealed. \emph{Astrophys. J.} \textbf{2006}, \emph{647},
  L175\relax
\mciteBstWouldAddEndPuncttrue
\mciteSetBstMidEndSepPunct{\mcitedefaultmidpunct}
{\mcitedefaultendpunct}{\mcitedefaultseppunct}\relax
\EndOfBibitem
\bibitem[Freindorf \latin{et~al.}(2021)Freindorf, Beiranvand, Delgado, Tao, and
  Kraka]{freindorf2021formation}
Freindorf,~M.; Beiranvand,~N.; Delgado,~A.~A.; Tao,~Y.; Kraka,~E. On the
  formation of CN bonds in Titan’s atmosphere—a unified reaction valley
  approach study. \emph{J. Mol. Model.} \textbf{2021}, \emph{27}, 1--20\relax
\mciteBstWouldAddEndPuncttrue
\mciteSetBstMidEndSepPunct{\mcitedefaultmidpunct}
{\mcitedefaultendpunct}{\mcitedefaultseppunct}\relax
\EndOfBibitem
\bibitem[Cable \latin{et~al.}(2012)Cable, H{\"o}rst, Hodyss, Beauchamp, Smith,
  and Willis]{Cable2012}
Cable,~M.~L.; H{\"o}rst,~S.~M.; Hodyss,~R.; Beauchamp,~P.~M.; Smith,~M.~A.;
  Willis,~P.~A. Titan tholins: simulating Titan organic chemistry in the
  Cassini-Huygens era. \emph{Chem. Rev.} \textbf{2012}, \emph{112},
  1882--1909\relax
\mciteBstWouldAddEndPuncttrue
\mciteSetBstMidEndSepPunct{\mcitedefaultmidpunct}
{\mcitedefaultendpunct}{\mcitedefaultseppunct}\relax
\EndOfBibitem
\bibitem[Skouteris \latin{et~al.}(2015)Skouteris, Balucani, Faginas-Lago,
  Falcinelli, and Rosi]{skouteris2015dimerization}
Skouteris,~D.; Balucani,~N.; Faginas-Lago,~N.; Falcinelli,~S.; Rosi,~M.
  Dimerization of methanimine and its charged species in the atmosphere of
  Titan and interstellar/cometary ice analogs. \emph{Astron. Astrophys.}
  \textbf{2015}, \emph{584}, A76\relax
\mciteBstWouldAddEndPuncttrue
\mciteSetBstMidEndSepPunct{\mcitedefaultmidpunct}
{\mcitedefaultendpunct}{\mcitedefaultseppunct}\relax
\EndOfBibitem
\bibitem[Granovsky(2011)]{xmcqdpt2}
Granovsky,~A. Extended multi-configuration quasi-degenerate perturbation
  theory: The new approach to multi-state multi-reference perturbation theory.
  \emph{J. Chem. Phys.} \textbf{2011}, \emph{134}, 214113\relax
\mciteBstWouldAddEndPuncttrue
\mciteSetBstMidEndSepPunct{\mcitedefaultmidpunct}
{\mcitedefaultendpunct}{\mcitedefaultseppunct}\relax
\EndOfBibitem
\bibitem[Granovsky()]{Firefly}
Granovsky,~A.~A. Firefly version 8. \url{http://classic.chem.msu.su}, Accessed
  on October 14, 2025\relax
\mciteBstWouldAddEndPuncttrue
\mciteSetBstMidEndSepPunct{\mcitedefaultmidpunct}
{\mcitedefaultendpunct}{\mcitedefaultseppunct}\relax
\EndOfBibitem
\bibitem[Zhu \latin{et~al.}(2019)Zhu, Thompson, and
  Mart{\'\i}nez]{zhu2019geodesic}
Zhu,~X.; Thompson,~K.~C.; Mart{\'\i}nez,~T.~J. Geodesic interpolation for
  reaction pathways. \emph{J. Chem. Phys.} \textbf{2019}, \emph{150}\relax
\mciteBstWouldAddEndPuncttrue
\mciteSetBstMidEndSepPunct{\mcitedefaultmidpunct}
{\mcitedefaultendpunct}{\mcitedefaultseppunct}\relax
\EndOfBibitem
\bibitem[Domcke \latin{et~al.}(2004)Domcke, Yarkony, and Köppel]{book:CI}
Domcke,~W.; Yarkony,~D.~R.; Köppel,~H. \emph{Conical Intersections}; WORLD
  SCIENTIFIC, 2004\relax
\mciteBstWouldAddEndPuncttrue
\mciteSetBstMidEndSepPunct{\mcitedefaultmidpunct}
{\mcitedefaultendpunct}{\mcitedefaultseppunct}\relax
\EndOfBibitem
\bibitem[Tully(1990)]{tully1990molecular}
Tully,~J.~C. Molecular dynamics with electronic transitions. \emph{J. Chem.
  Phys.} \textbf{1990}, \emph{93}, 1061--1071\relax
\mciteBstWouldAddEndPuncttrue
\mciteSetBstMidEndSepPunct{\mcitedefaultmidpunct}
{\mcitedefaultendpunct}{\mcitedefaultseppunct}\relax
\EndOfBibitem
\bibitem[Granucci \latin{et~al.}(2010)Granucci, Persico, and
  Zoccante]{Granucci2010}
Granucci,~G.; Persico,~M.; Zoccante,~A. Including quantum decoherence in
  surface hopping. \emph{J. Chem. Phys.} \textbf{2010},
  \emph{133}, 134111\relax
\mciteBstWouldAddEndPuncttrue
\mciteSetBstMidEndSepPunct{\mcitedefaultmidpunct}
{\mcitedefaultendpunct}{\mcitedefaultseppunct}\relax
\EndOfBibitem
\bibitem[Belyaev and Lebedev(2011)Belyaev, and Lebedev]{PhysRevA.84.014701}
Belyaev,~A.~K.; Lebedev,~O.~V. Nonadiabatic nuclear dynamics of atomic
  collisions based on branching classical trajectories. \emph{Phys. Rev. A}
  \textbf{2011}, \emph{84}, 014701\relax
\mciteBstWouldAddEndPuncttrue
\mciteSetBstMidEndSepPunct{\mcitedefaultmidpunct}
{\mcitedefaultendpunct}{\mcitedefaultseppunct}\relax
\EndOfBibitem
\bibitem[Varga \latin{et~al.}(2018)Varga, Parker, and Truhlar]{varga2018direct}
Varga,~Z.; Parker,~K.~A.; Truhlar,~D.~G. Direct diabatization based on
  nonadiabatic couplings: the N/D method. \emph{Phys. Chem. Chem. Phys.}
  \textbf{2018}, \emph{20}, 26643--26659\relax
\mciteBstWouldAddEndPuncttrue
\mciteSetBstMidEndSepPunct{\mcitedefaultmidpunct}
{\mcitedefaultendpunct}{\mcitedefaultseppunct}\relax
\EndOfBibitem
\bibitem[Li \latin{et~al.}(2021)Li, Liu, Li, Zhai, Yang, Qu, and Li]{li2021new}
Li,~Y.; Liu,~J.; Li,~J.; Zhai,~Y.; Yang,~J.; Qu,~Z.; Li,~H. A new
  permutation-symmetry-adapted machine learning diabatization procedure and its
  application in MgH{$_2$} system. \emph{J. Chem. Phys.} \textbf{2021},
  \emph{155}, 214102\relax
\mciteBstWouldAddEndPuncttrue
\mciteSetBstMidEndSepPunct{\mcitedefaultmidpunct}
{\mcitedefaultendpunct}{\mcitedefaultseppunct}\relax
\EndOfBibitem
\bibitem[Richardson(2023)]{richardson2023machine}
Richardson,~J.~O. Machine learning of double-valued nonadiabatic coupling
  vectors around conical intersections. \emph{J. Chem. Phys.} \textbf{2023},
  \emph{158}, 011102\relax
\mciteBstWouldAddEndPuncttrue
\mciteSetBstMidEndSepPunct{\mcitedefaultmidpunct}
{\mcitedefaultendpunct}{\mcitedefaultseppunct}\relax
\EndOfBibitem
\bibitem[Zhang \latin{et~al.}(2024)Zhang, Pios, Martyka, Ge, Hou, Chen, Chen,
  Jankowska, Barbatti, and Dral]{MLatom_paper}
Zhang,~L.; Pios,~S.~V.; Martyka,~M.; Ge,~F.; Hou,~Y.-F.; Chen,~Y.; Chen,~L.;
  Jankowska,~J.; Barbatti,~M.; Dral,~P.~O. MLAtom software ecosystem for
  surface hopping dynamics in Python with quantum mechanical and machine
  learning methods. \emph{J. Chem. Theo. Comput.} \textbf{2024}, \emph{20},
  5043--5057\relax
\mciteBstWouldAddEndPuncttrue
\mciteSetBstMidEndSepPunct{\mcitedefaultmidpunct}
{\mcitedefaultendpunct}{\mcitedefaultseppunct}\relax
\EndOfBibitem
\bibitem[Batatia \latin{et~al.}(2022)Batatia, Kovacs, Simm, Ortner, and
  Cs{\'a}nyi]{batatia2022mace}
Batatia,~I.; Kovacs,~D.~P.; Simm,~G.; Ortner,~C.; Cs{\'a}nyi,~G. MACE: Higher
  order equivariant message passing neural networks for fast and accurate force
  fields. \emph{Adv. Neural Inf. Process Syst.} \textbf{2022}, \emph{35},
  11423--11436\relax
\mciteBstWouldAddEndPuncttrue
\mciteSetBstMidEndSepPunct{\mcitedefaultmidpunct}
{\mcitedefaultendpunct}{\mcitedefaultseppunct}\relax
\EndOfBibitem
\end{mcitethebibliography}

\providecommand{\latin}[1]{#1}
\makeatletter
\providecommand{\doi}
  {\begingroup\let\do\@makeother\dospecials
  \catcode`\{=1 \catcode`\}=2 \doi@aux}
\providecommand{\doi@aux}[1]{\endgroup\texttt{#1}}
\makeatother
\providecommand*\mcitethebibliography{\thebibliography}
\csname @ifundefined\endcsname{endmcitethebibliography}
  {\let\endmcitethebibliography\endthebibliography}{}

\end{document}